\documentclass[fp,twocolumn]{jpsj3}
\usepackage{txfonts}
\title{High-Field Fermi Surface Properties in the Low Carrier Heavy Fermion Compound URu$_2$Si$_2$}

\author{
Dai~{\sc Aoki}$^{1,2}$\thanks{E-mail address: dai.aoki@cea.fr}, 
Georg~{\sc Knebel}$^1$\thanks{E-mail address: georg.knebel@cea.fr},
Ilya~{\sc Sheikin}$^3$,
Elena~{\sc Hassinger}$^1$,
Liam~{\sc Malone}$^4$
Tatsuma~D.~{\sc Matsuda}$^{5,1}$, 
and
Jacques~{\sc Flouquet}$^1$
}

\inst{%
$^1$INAC, SPSMS, CEA-Grenoble, 38054 Grenoble, France\\
$^2$IMR, Tohoku University, Oarai, Ibaraki 311-1313, Japan\\
$^3$LNCMI-G, CNRS, 38042 Grenoble, France\\
$^4$LNCMI-T, UPR 3228 (CNRS-INSA-UJF-UPS), Toulouse 31400, France\\
$^5$ASRC, JAEA, Tokai, Ibaraki 319-1195, Japan\\ 
}

\abst{
We performed the Shubnikov-de Haas (SdH) experiments of the low carrier heavy fermion compound URu$_2$Si$_2$
at high fields up to $34\,{\rm T}$ and at low temperatures down to $30\,{\rm mK}$.
All main SdH branches named $\alpha$, $\beta$ and $\gamma$ were observed for all the measured field-directions
($H \parallel [001]\to [100]$, $[100] \to [110]$ and $[001] \to [110]$),
indicating that these are attributed to the closed Fermi surfaces with nearly spherical shapes.
Anomalous split of branch $\alpha$ was detected for the field along the basal plane,
and the split immediately disappears by tilting the field to $[001]$ direction,
implying a fingerprint of the hidden order state.
High field experiments reveal the complicated field-dependence of the SdH frequencies and the cyclotron masses
due to the Zeeman spin-splitting associated with the Fermi surface reconstruction in the hidden order state with small carrier numbers.
A new SdH branch named $\omega$ with large cyclotron mass of $25\,m_0$ was detected at high fields above $23\,{\rm T}$
close to the hidden order instabilities.
}

\kword{Shubnikov-de Haas effect, de Haas-van Alphen effect, Fermi surface, cyclotron effective mass, spin split, hidden order, URu$_2$Si$_2$}

\begin{document}
\maketitle
\section{Introduction}
Quantum oscillation measurements, such as de Haas-van Alphen (dHvA) effect and Shubnikov-de Haas (SdH) effect, 
are one of the most powerful experimental probes to study the Fermi surface properties from a microscopic point of view.~\cite{Sho84,Onu95a}
The heavy quasi-particle band with small Fermi surface is easily affected by the strong magnetic field,
consequently the unusual spin-splitting associated with the Fermi surface reconstruction,
which may lead to a Lifshitz transition~\cite{Lif60}, can be precisely detected by the dHvA or SdH experiments.

The well-known low carrier heavy fermion compound URu$_2$Si$_2$
is one of the ideal system to study the Fermi surface reconstruction due to the magnetic field.
It crystallizes in the ThCr$_2$Si$_2$-type tetragonal structure (space group: $I4/mmm$, No.~139).
Below $T_0 = 17.5\,{\rm K}$ the so-called ``hidden order'' (HO) phase emerges,
where the carrier number is reduced to be one-third compared to that of the paramagnetic phase.
An extrinsic ordered moment which is less than $0.05\,\mu_{\rm B}$ with the antiferromagnetic $q$-vector $Q_{\rm AF}=(0,0,1)$
is detected by neutron scattering experiments.~\cite{Bro87,Bou03,Wie07}
A key fingerprint of the HO state is the emergence of a magnetic resonance at $Q_0=(1,0,0)$ 
with an energy $E_0\sim 2\,{\rm meV}$.
Applying pressure, this resonance disappears when the antiferromagnetic phase is established.~\cite{Vil08}
Interestingly, the superconducting phase ($T_{\rm c}\sim 1.5\,{\rm K}$ at ambient pressure) is embedded in the HO phase.
It exhibits a highly anisotropic upper critical field at $0\,{\rm K}$, $H_{\rm c2}\sim 14\,{\rm T}$ for $H\parallel [100]$ and $H_{\rm c2}\sim 3\,{\rm T}$ for $H\parallel [001]$
associated with a large anisotropy of the initial slope $\partial H_{\rm c2}/\partial T$ at $H=0$.
Such a large anisotropic $\partial H_{\rm c2}/\partial T$ cannot be explained by the observed Fermi surfaces
which are almost spherical without a large anisotropic effective masses~\cite{Ohk99,Bri95}.

Many theoretical scenarios including various rank multipole order have been proposed to identify the HO phase, 
but up to date there are no definite experimental conclusions for the order parameter of HO phase.
However, the experimental proofs and theoretical interpretations with multipole model
are still unclear. 
Recent torque measurements using small and high quality single crystals propose a breaking of the four-fold symmetry in the basal plane 
below $T_{\rm 0}$, suggesting a ``Nematic'' state~\cite{Oka11}. 

The Fermi surface study by quantum oscillation measurements, which have been established as a microscopic probe with high accuracy,
is a key in order to investigate the electronic state associated with the HO state.
According to previous dHvA experiments at ambient pressure~\cite{Ohk99}, 
three kinds of Fermi surface sheets ($\alpha$, $\beta$ and $\gamma$) were detected from $H \parallel [001]$ to $[100]$, 
while only branch $\alpha$ was observed from $H\parallel  [100]$ to $[110]$.
Our recent SdH experiments clarified that branch $\beta$ is four-folded in the simple tetragonal Brillouin zone~\cite{Has10_URu2Si2},
and is located between $\Gamma$ point and $X$ point from the comparison to the band calculation~\cite{Opp10,Harima_pub,Suzuki_pub}.
The frequencies of branch $\alpha$ as well as $\beta$ and $\gamma$ do not change significantly even in the AF state,
despite the fact that the cyclotron masses gradually decreases with pressure,
indicating that the Fermi surfaces in the AF state are rather similar to those in the HO state\cite{Has10_URu2Si2,Nak03}.
The Brillouin zone in the HO state is based on the simple tetragonal symmetry, which 
is the same as that in the AF state with $Q_{\rm AF}=(0,0,1)$.

Here we report the high field SdH experiments up to $34\,{\rm T}$ close to the spin polarized phase
in order to clarify the Fermi surface properties and the instabilities in the HO state.
We observed an unusual Zeeman spin-splitting and a Fermi surface reconstruction directly by the SdH experiments.

\section{Experimental}
High quality single crystals of URu$_2$Si$_2$ were grown by the Czochralski method using a tetra-arc furnace.
The details are described in Ref.~\citen{Aok10}.
We performed the SdH experiments using the same single crystal, which have been used in our previous SdH experiments,
where the SdH signal starts to appear at low field ($\approx 3.5\,{\rm T}$).
The residual resistivity ratio (RRR) is estimated to be more than $500$,
indicating the very high quality of our sample.
The detail studies on transport properties depending on the sample qualities are recently reported in ref.~\citen{Mat11}.
High field SdH experiments were done using a top-loading dilution refrigerator with a sample rotation mechanism 
at low temperatures down to $30\,{\rm mK}$ and at high fields up to $34\,{\rm T}$ by a resistive magnet at LNCMI-Grenoble.
The samples were rotated for the field direction from $[001]$ to $[100]$, and from $[100]$ to $[110]$
in the tetragonal structure.
The SdH signal was measured by the four-probe AC method with the electrical current $J$ along $[100]$ direction ($J \sim 100\,\mu{\rm A}$, frequency: $\sim 30\,{\rm Hz}$).
A conventional dilution refrigerator with a low temperature transformer was also utilized for the low field measurements
below $13\,{\rm T}$ for the field direction from $[001]$ to $[100]$, and from $[001]$ to $[110]$.
To clarify the split of branch $\alpha$ along $H \parallel [100]$, 
dHvA experiments were also performed by field modulation method at fields up to $15\,{\rm T}$ and at low temperatures down to $20\,{\rm mK}$. 
The great advantage compared to the SdH experiments is that one can detect branch $\alpha$ at low fields even in the superconducting mixed state~\cite{Onu99}.

\section{Results and Discussion}
\begin{figure}[tbh]
\begin{center}
\includegraphics[width=1 \hsize,clip]{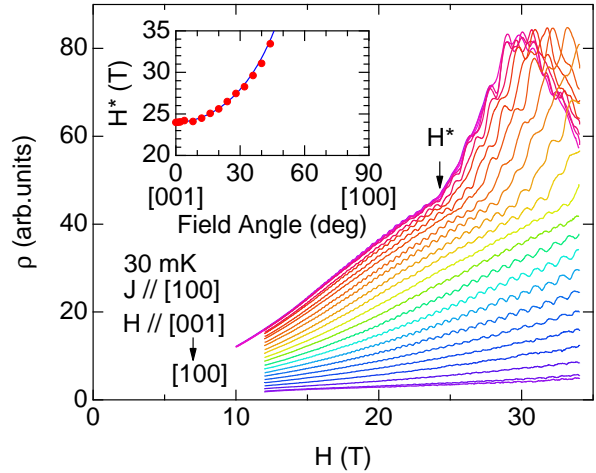}
\end{center}
\caption{(Color online) Magnetoresistance at $30\,{\rm mK}$ for $J \parallel [100]$ at different field directions from $[001]$ to $[100]$ in URu$_2$Si2$_2$. The current was applied in $[100]$ direction. The inset shows the angular dependence of $H^\ast$ at which the magnetoresistance indicates a kink. The solid line indicates the $1/\cos\theta$ dependence. Note that the magnetoresistance decreases with field angle, 
because it changes from transverse to longitudinal magnetoresistance.
}
\label{fig:MR_AngDep}
\end{figure}

Figure~\ref{fig:MR_AngDep} shows the magnetoresistance $\rho (H)$ at different field angles from $[001]$ to $[100]$ at $30\,{\rm mK}$. Very clear SdH oscillations can be seen in the raw data of magnetoresistance, demonstrating the high quality of the present sample.
For $H \parallel [001]$, the transverse magnetoresistance increases with $H^2$ dependence up to $\sim 15\,{\rm T}$
and shows a broad convex curvature at higher fields.
Further increasing field, a kink appears at $H^\ast=24\,{\rm T}$.
Above $H^\ast$ the magnetoresistance rapidly increases, revealing a maximum at $\sim 29\,{\rm T}$ and then decreases.
Rotating the field angle $\theta$ from $[001]$ to $[100]$, 
$H^\ast$ increases as a function of $1/\cos\theta$, as shown in the inset of Fig.~\ref{fig:MR_AngDep}.
These results are in good agreement with those in the previous reports \cite{Jo2007,Shi09,Altarawneh2011,Sch12}, 
although the value of $H^\ast$ in ref.~\citen{Shi09} ($H^\ast =22.5\,{\rm T}$) is slightly lower than the present data.

\begin{figure}[tbh]
\begin{center}
\includegraphics[width=1 \hsize,clip]{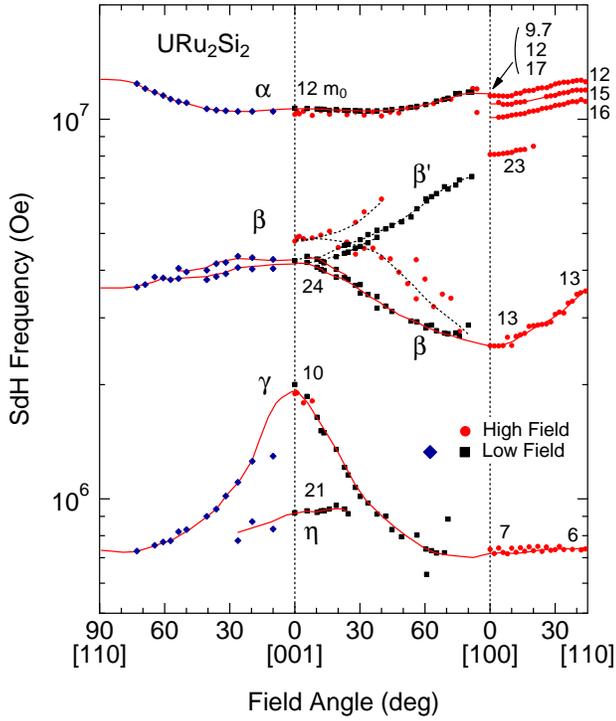}
\end{center}
\caption{Angular dependence of the SdH frequency in URu$_2$Si$_2$. Red circles are the results at high field. Dark blue diamonds are the results at low field below $13\,{\rm T}$. Black square are our previous results at low field.~\protect\cite{Has10_URu2Si2}
Branch $\beta$ at high fields between $H\parallel [001]$ and $[100]$ is shifted because of the field-dependent frequency.
Cyclotron effective masses are also shown in numbers for $H\parallel [001]$, $[100]$ and $[110]$.}
\label{fig:Freq_AngDep}
\end{figure}

Figure~\ref{fig:Freq_AngDep} shows the angular dependence of the SdH frequency.
Branch $\alpha$ is almost constant against all the field angle, indicating a nearly spherical Fermi surface.
Branch $\beta$ splits for $H\parallel [001]$ to $[100]$ while it degenerates for $H \parallel [001]$ to $[110]$,
indicating the four-fold Fermi surface.
Branch $\gamma$ originates from the small ellipsoidal Fermi surface squeezed along $[001]$ direction.

The frequency of branch $\beta$ at high field guided by doted line is slightly shifted to higher frequency,
compared to that at low field.
This is due to the field dependence of the frequency, as discussed later.
We did not observe any signatures of branch $\varepsilon$ ($F \sim 1.5 \times 10^7\,{\rm Oe}$ for $H\parallel c$-axis) which have been reported in ref.~\citen{Shi09}.
\begin{figure}[tbh]
\begin{center}
\includegraphics[width=0.5 \hsize,clip]{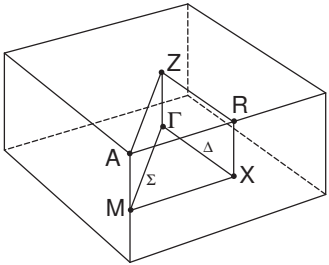}
\end{center}
\caption{Brillouin zone of the simple tetragonal structure.}
\label{fig:BZ}
\end{figure}

In the previous report~\cite{Ohk99} only branch $\alpha$ was detected in the basal plane from $H\parallel [100]$ to $[110]$, 
while we detected all the main branches $\alpha$, $\beta$ and $\gamma$ in the present experiments.
The angular dependence of branch $\beta$ and $\beta^\prime$ is consistent with our previous results \cite{Has10_URu2Si2},
in good agreement with the results of band calculation,
where the four-folded Fermi surface is expected \cite{Opp10,Harima_pub,Suzuki_pub}. 
Branch $\beta^\prime$ was not observed in the basal plane probably due to the unfavorable curvature factor.
A striking point is that branch $\alpha$ split into three frequencies, which is also consistent with the previous report.~\cite{Ohk99}

We have determined the cyclotron effective mass $m_{\rm c}^\ast$
from the temperature dependence of the SdH amplitude
for $H\parallel [001]$, $[100]$ and $[110]$ directions, as shown by annotations in numbers in Fig.~\ref{fig:Freq_AngDep}.
The obtained values for $H\parallel [001]$ are in good agreement with the previously reported values.
The cyclotron mass of branch $\gamma$ is slightly reduced to be $6$--$7\,m_0$ in the basal plane.
Branch $\beta$ also shows the decrease of the cyclotron mass which is $13\,m_0$ in the basal plane. 
Our previous low field ($H< 13$~T) measurements~\cite{Has10_URu2Si2} show that 
the mass of branch $\beta^\prime$ is almost isotropic, ranging from $25\,m_0$ to $30\,m_0$ for the field direction between $[001]$ and $[100]$.
Thus we believe the mass of $\beta^\prime$ in the basal plane is also similar,
although we could not detect $\beta^\prime$ in the basal plane.
Branch $\alpha$ indicates an isotropic cyclotron mass,
although the masses for splitting branch $\alpha$ are slightly increased for $H \parallel [110]$.

Before the assignment of detected SdH branches to the Fermi surfaces,
we roughly evaluate the Sommerfeld coefficient ($\gamma$-value) deduced from the SdH experiments.
Assuming the spherical Fermi surfaces,~\cite{Aok10}
the calculated $\gamma$-values for branches $\alpha$, $\beta$ and $\gamma$ are 
$\gamma_{\alpha} \approx 6.1\,{\rm mJ/(K^2 mol)}$,
$\gamma_{\beta}  \approx 7.8\,{\rm mJ/(K^2 mol)}$ and
$\gamma_{\gamma} \approx 2.7\,{\rm mJ/(K^2 mol)}$, respectively.
Since branch $\beta$ is attributed to four-fold Fermi surfaces and no two-fold Fermi surfaces is predicted in the 
band calculations, that is no Fermi surfaces centered at X point,
the total $\gamma$-value by SdH experiments is 
$\gamma_{\rm SdH}^{} = \gamma_{\alpha} + 4\gamma_{\beta} + \gamma_\gamma \approx 40\,{\rm mJ/(K^2 mol)}$,
while $\gamma$-value by the specific heat measurement is known to be $\gamma_{\rm Cp}^{}=55\,{\rm mJ/(K^2 mol)}$,
Thus the $\gamma$-value from an undetected Fermi surface should be $15\,{\rm mJ/(K^2 mol)}$,
meaning that $30\,{\%}$ of the total $\gamma$-value is still missing.

The SdH results indicate that there are three kinds of closed Fermi surfaces in the Brillouin zone based on the simple tetragonal symmetry.
According to the recent band calculations,~\cite{Elg09,Har09,Suzuki_pub} 
two kinds of electron Fermi surfaces (e-FS) centered at $M$ point 
(see Fig.~\ref{fig:BZ} for the simple tetragonal Brillouin zone),
four-fold symmetry e-FS located between $\Gamma$ and $X$ point (namely on $\Delta$ line)
and hole Fermi surface (h-FS) centered at $\Gamma$ point 
are predicted as main Fermi surfaces. 
The h-FS has a slightly complicated shape accompanying with a small wing-shaped Fermi surfaces located between $\Gamma$ and $M$ point.
This wing-shaped FS can be easily affected in topology between HO and AF state.

On the basis of these band calculations,
two different possibilities for the assignment of SdH branches can be considered:

\begin{description}
\item[(i)] Branch $\alpha$ is attributed to the h-FS centered at $\Gamma$ point. 
Branch $\beta$ is attributed to the four-fold e-FS located between $\Gamma$ and $X$ point.
Branch $\gamma$ arises from e-FS at $M$ point. 
The $\eta$ branch is assigned to the small FS at the $Z$ point.
In this case, a missing Fermi surface is the e-FS centered at $M$ point. 
\item[(ii)]
Branches $\alpha$ and $\gamma$ are attributed to two different e-FS at $M$ point.
Branch $\beta$ is attributed to the four-fold e-FS located between $\Gamma$ and $X$ point.
In this case, a missing Fermi surface is the h-FS centered at $\Gamma$ point. 
This assignment corresponds to that in ref.~\citen{Opp10} and is based on the observation of a rather light $\varepsilon$ branch in ref.~\citen{Shi09} which increases along the $[100]$ direction. 
However, as already mentioned above, in our experiment we do not see any indications of this frequency, and further it is unclear whether this orbit would be a fundamental orbit as it has been reported only for $H > 17$~T and, as will be discussed below, strong field induced modifications of the FS appear.
\end{description}

Although we cannot definitely conclude from our experiments which case is correct, 
the case (i) seems to be more plausible.
The recent band calculations predict that the 5$f$-electron contribution to the Fermi surface
is dependent on the bands~\cite{Suzuki_pub}, 
and the e-FS has more 5$f$-contribution than h-FS,
indicating the existence of heavy e-FS.
This is consistent with our results where the light h-FS corresponds to branch $\alpha$
and the heavy e-FS corresponds to branch $\beta$.
Since the $f$-component of e-FS at $M$ point is similar to or even larger than that of branch $\beta$,
it is expected that the missing FS has also large cyclotron mass which might be more than $30\,m_0$.
In our experiment, the fact that we fail to observe this missing FS can be 
the combined effects of large cyclotron mass and unfavorable curvature factor.
The assignment of case (i) is also consistent with the results of thermal conductivity measurements~\cite{Kas07},
in which a heavy e-FS is predicted.

An important question is the relation between the Fermi surfaces and the anisotropy of the superconducting upper critical field $H_{\rm c2}$
as well as anisotropic initial slope of $H_{\rm c2}$.
All the detected Fermi surfaces are isotropic, and the cyclotron masses are also rather isotropic.
A plausible idea would be the existence of an ellipsoidal Fermi surface elongated along $[001]$ direction
with hot spots, 
where the $f$-electron dominantly contributes to the conduction band forming 
a heavy band.
The emergence of hot spots on the Fermi surface associated with the antiferromagnetic fluctuations and 
the increase of effective mass are discussed already in CeIn$_3$.~\cite{Gor06}

\begin{figure}[tbh]
\begin{center}
\includegraphics[width=1 \hsize,clip]{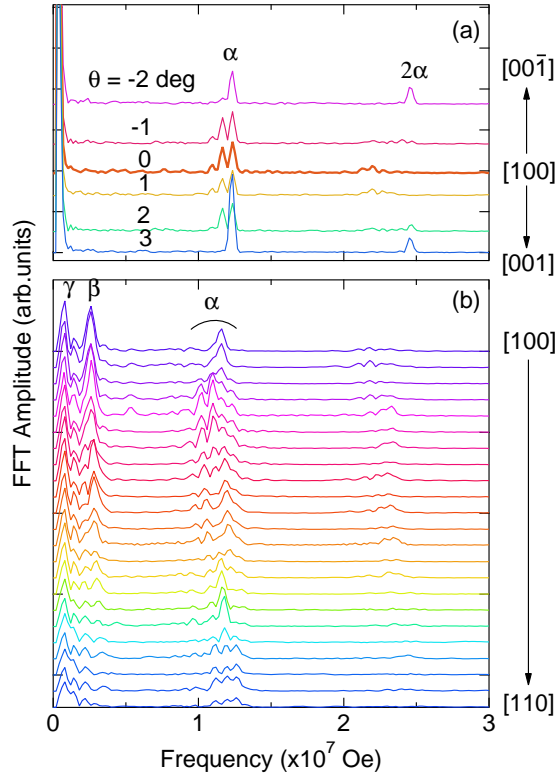}
\end{center}
\caption{(a) FFT spectra of the dHvA oscillations in the field range from $8$ to $15\,{\rm T}$ for the field direction close to $[100]$. $\theta$ is a tilt angle from $[100]$ to $[001]$ direction. (b) FFT spectra of the SdH oscillations above $15\,{\rm T}$ for the field directions from $[100]$ to $[110]$.}
\label{fig:Alpha_split}
\end{figure}

Interestingly, the single FFT peak of branch $\alpha$ splits into three frequencies in the basal plane
($\alpha_1\sim 1.24\times 10^7\,{\rm Oe}$, 
$\alpha_2\sim 1.16\times 10^7\,{\rm Oe}$ and 
$\alpha_3\sim 1.09\times 10^7\,{\rm Oe}$ for $H\parallel [100]$). 
Figure~\ref{fig:Alpha_split}(a) shows the FFT spectra of the dHvA oscillations near $H\parallel [100]$.
When the field is slightly tilted more than $3\,{\rm deg}$ from $[100]$ to $[001]$, the split disappears immediately
and branch $\alpha_1$ remains as an original frequency.
On the other hand, the splitting is always observed for the field along the basal plane, retaining almost the same splitting width ($\Delta F$) 
as shown in Figs.~\ref{fig:Freq_AngDep} and \ref{fig:Alpha_split}(b).
It should be noted that in Fig.~\ref{fig:Alpha_split}(b) many FFT peaks more than three are observed at some field angles 
because of the sum or subtraction with branch $\gamma$.
The similar splitting of branch $\alpha$ is also reported in the previous paper.~\cite{Ohk99}
As seen in Fig.~\ref{fig:Alpha_split}(b), 
the relative amplitude of FFT between $\alpha_1$, $\alpha_2$ and $\alpha_3$
also shows an unusual angular dependence.
For example, the amplitude of branch $\alpha_1$ is dominant for $H\parallel [100]$, but
the amplitude of branch $\alpha_2$ becomes dominant at $42\,{\rm deg}$ tilted from $[100]$.
The splitting has no significant field-dependence at least from $8$ to $15\,{\rm T}$,
which was confirmed by the dHvA experiments for $H\parallel [100]$.

In general, if the Fermi surface is corrugated with maximal and minimal cross-sectional area, a splitting of the Fermi surface branch can be detected.
However it is improbable for branch $\alpha$, because the split is observed all in the basal plane with the same splitting width in frequencies.
Furthermore, the band calculations do not predict such a corrugated Fermi surface.

When the magnetization has a non-linear response such as (pseudo-)metamagnetism, one can observe the split of frequency.
In this case, two frequencies originating from up-spin and down-spin Fermi surfaces will be observed,
and the observed frequencies which correspond to the back projection to zero field will show a field dependence.
We can exclude such a simple spin-split effect for the split of branch $\alpha$,
because three frequencies are observed and they are not field-dependent.

An interesting scenario is a breaking of four-fold symmetry under magnetic fields
when the field is applied parallel to the basal plane.
Because of the strong magnetic field and the corresponding magnetostriction effect, 
the tetragonal symmetry is changed into the orthogonal symmetry.
This effect is usually so small that one can observe only the Fermi surfaces based on the tetragonal symmetry.
However, if the effect is large for some reason, the frequency is shifted from the original one
and a mono-domain state is formed.
By rotating the field angle, a switching of domain from one to the other occurs at certain field angle.
Correspondingly the frequency is abruptly changed or splits.
This is in fact observed in CeSb with cubic symmetry, where the three-fold ellipsoidal Fermi surface exists in the bcc Brillouin zone~\cite{Cra87}.
This domain switching effect seems to agree with the fact that FFT amplitude of branch $\alpha$ abruptly changes at certain field angle as shown in Fig.~\ref{fig:Alpha_split}(b).
Since the split of branch $\alpha$ is always observed in the basal plane, 
the perfect monomodain is not formed, but the multi-domain is probably formed effectively.

\begin{figure}[tbh]
\begin{center}
\includegraphics[width=0.6 \hsize,clip]{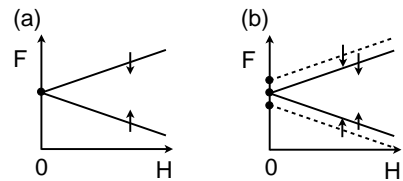}
\end{center}
\caption{Relation between dHvA frequency for up/down-spin Fermi surfaces and magnetic field is schematically shown (a) in the usual case, and (b) in the present case. Closed circles at zero field correspond to the observed frequencies.}
\label{fig:Split_schem}
\end{figure}

The reason why three frequencies are observed might be explained as follows.
The observed dHvA frequency corresponds to the back projection to zero field, and 
the observed frequency for up-spin Fermi surfaces due to the Zeeman splitting coincides with that for down-spin Fermi surface (see Fig.~\ref{fig:Split_schem}(a)).
In the present case, an original Fermi surface, up-spin and down-spin Fermi surfaces are observed simultaneously because of the formation of the multi-domain, as shown in Fig.~\ref{fig:Split_schem}(b).

The reason for the formation of the multi-domain structure is not clear. 
It is worth noting that the breaking of four-fold symmetry was reported below $T_0$ by magnetic torque measurements using high quality and very small samples~\cite{Oka11}.
Recently, a model of rank-5 multipole order with nematic symmetry is proposed to explain the breaking of four-fold symmetry.~\cite{Ike12}
In this calculation, the significant difference of Fermi surfaces between HO state and AF state is:
i) the breaking of four-fold symmetry may occur on the cage-like Fermi surface located between $\Gamma$ and $M$ point in the HO state
ii) two Fermi surfaces centered at M point may be strongly affected.
In our experiments, we cannot conclude if the split of branch $\alpha$ is related with the symmetry breaking and the hidden order.
However, the effective internal field probably induces the split of branch $\alpha$, as shown in Fig.~\ref{fig:Split_schem}(b).
Interesting open questions are: i) Why does the effective internal field appear? ii) Does the split of branch $\alpha$ merge into single branch in the AF state?
We believe the splitting of branch $\alpha$, which can be explained neither by simple Zeeman-splitting nor by the corrugation of FS, is a fingerprint of HO state.
The key experiments for the future will be precise SdH or dHvA studies under pressure with fine-tuning of the field direction.

Another possible explanation for the splitting of branch $\alpha$ is the breaking of inversion symmetry under magnetic field
and the magnetic breakdown between two splitting Fermi surfaces.
This is in fact observed in non-centrosymmetric compounds CeRhSi$_3$ and LaRhSi$_3$.~\cite{Ter08}

Finally. let us now focus on the field effect in the SdH experiments which can already be observed at rather low fields. 
Previously topological changes of the Fermi surface under high field have been reported on the basis of SdH and thermoelectric power experiments~\cite{Shi09,Altarawneh2011,Malone2011}. 
As discussed above, ref.~\citen{Shi09} reports the observation of a new SdH frequency $\varepsilon$ for $H > 17$--$20\,{\rm T}$ in Hall resistivity measurements.
Here in transverse magnetoresistivity measurements we do not see any signature of this frequency. 
From the thermopower experiments however, topological changes of the Fermi surface at fields of $H\approx 12\,{\rm T}$ and $H = H^\ast \approx 23\,{\rm T}$ could be evidenced~\cite{Malone2011}. 
A cascade of field-induced Fermi surface changes had furthermore been discussed in ref.~\citen{Altarawneh2011} which goes along with the interpretation of the thermopower experiments.
Clearly in this complex multi-band system, 
the modification on one FS gives rise to a feedback on the whole FS sheets with different consequences on the observed probes.

\begin{figure}[tbh]
\begin{center}
\includegraphics[width=1 \hsize,clip]{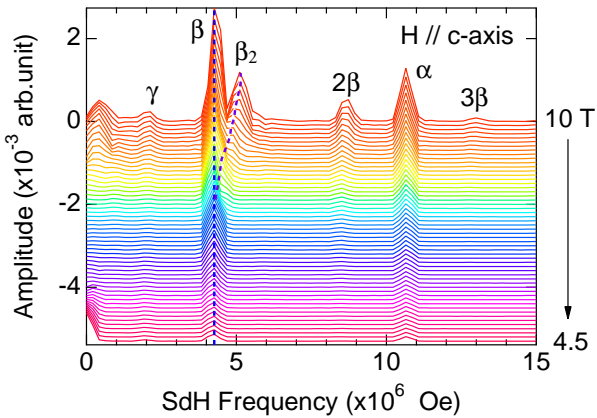}
\end{center}
\caption{(Color online) FFT spectra at different field ranges from $10$ to $4.5\,{\rm T}$ for $H \parallel [001]$.}
\label{fig:Beta_split}
\end{figure}

Even at rather low field the strong effect of the polarization of the Fermi surface pockets gets obvious.
Figure~\ref{fig:Beta_split} presents the FFT spectra for $H\parallel [001]$ at different field ranges from $4.5$ to $10\,{\rm T}$. Clearly, branch $\beta$ exhibits a splitting above $8\,{\rm T}$.
The frequency of the new split branch $\beta_2$ increases gradually with field,
while the frequency of the original main branch $\beta$ is almost constant or increases very slightly with field.
Branch $\beta_2$ is most likely due to the spin split. 
Interestingly, the cyclotron mass of branch $\beta_2$ is quite large ($40\,m_0$).
This value is almost twice as large as that of original branch $\beta$ ($24\,m_0$).
The spin dependent masses are observed in several heavy fermion compounds, such as CeRu$_2$Si$_2$~\cite{Taka96}, CePd$_2$Si$_2$~\cite{She03} and CeCoIn$_5$~\cite{Mcc05}. 
The other main branches $\alpha$, $\gamma$ does not show any significant changes of frequency at least up to $10\,{\rm T}$.

\begin{figure}[tbh]
\begin{center}
\includegraphics[width=1 \hsize,clip]{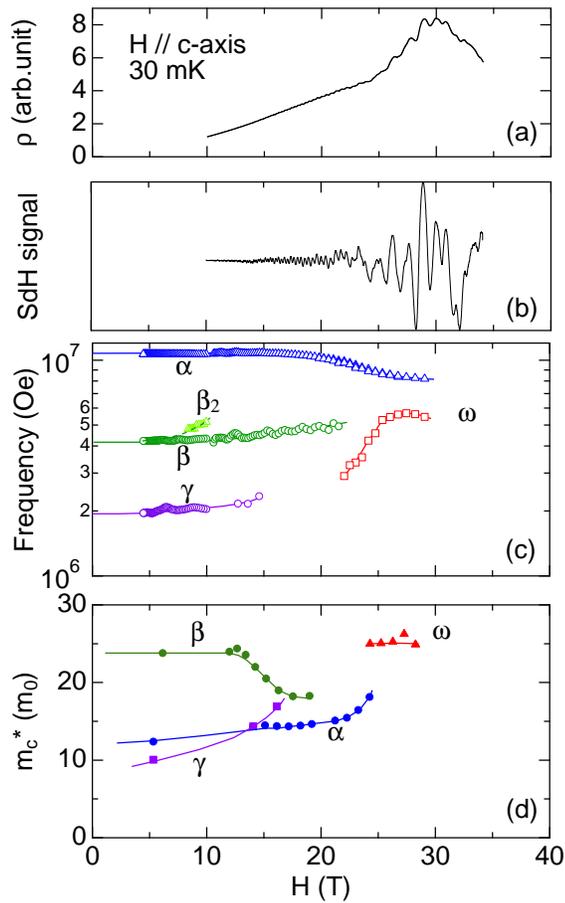}
\end{center}
\caption{(Color online) (a)Magnetoresistance for $H\parallel [001]$ at $30\,{\rm mK}$. (b)SdH signal obtained from the data in panel (a). (c)Field dependence of the SdH frequencies. (d)Field dependence of the cyclotron effective mass.}
\label{fig:Hdep_Freq_mass}
\end{figure}

The high field SdH experiments clearly demonstrate the field-dependence of the  SdH frequencies.
Figures~\ref{fig:Hdep_Freq_mass}(a) and \ref{fig:Hdep_Freq_mass}(b) represent the high field magnetoresistance up to $34\,{\rm T}$
and the corresponding SdH oscillations obtained by subtracting the non-oscillatory contribution.
Above $H^\ast = 24\,{\rm T}$, the amplitude of SdH oscillation is strongly enhanced,
and the frequencies seem to be drastically changed.

Figure~\ref{fig:Hdep_Freq_mass}(c) shows the field dependence of the SdH frequencies.
Frequencies were obtained by the FFT analysis, which are crosschecked by the maximum entropy method (MEM) 
in order to improve the frequency resolution.
Special attentions were payed to distinguish fundamental frequencies from the harmonics, sum and subtraction.

The frequency of branch $\alpha$ slightly decrease above $15\,{\rm T}$,
and shows the large decrease around $H^\ast$.
At $29\,{\rm T}$, it reaches $8.2\times 10^6\,{\rm Oe}$, which corresponds to more than $20\,\%$ drop from
the original frequency.
On the other hand, branch $\beta$ increases gradually with increasing field above $10\,{\rm T}$,
and reaches $5\times 10^6\,{\rm Oe}$ at $21\,{\rm T}$.
This is $20\,{\%}$ increase from the low field value.
The frequency of branch $\gamma$ increases with field as well.
A striking point is that a new SdH branch named $\omega$ with a large cyclotron mass appears approximately above $H^\ast$,
indicating the emergence of a new Fermi surface, which might be not only due to the Zeeman splitting but also related with a field-mixing of HO parameter.
The HO parameter here might be induced by the different multipole components
close to the vicinity of the multi-phase cascade domain on approaching the polarized paramagnetic phase.
The details of this new branch will be reported in another paper.

It should be noted that the observed frequency $F_{\rm obs}$ is not equal to
the true frequency $F_{\rm tr}$, because $F_{\rm obs}$ is the back projection to zero field.
Usually when the Zeeman spin-splitting energy increases linearly with field,
$F_{\rm obs}$ is constant against the field, as shown in Fig.~\ref{fig:Split_schem}(b)
However, if the Zeeman spin-splitting have non-linear field-response associated with the reconstruction of FS,
$F_{\rm obs}$ shows the field dependence.
When the magnetization increases with upward curvature such as (pseudo) metamagnetism,
the field-dependent $F_{\rm obs}$ is detected, as typically demonstrated in CeRu$_2$Si$_2$~\cite{Taka96} and UPt$_3$~\cite{Kim00}.
In URu$_2$Si$_2$, the magnetization increases linearly up to $30\,{\rm T}$
and then shows the slight upward curvature. At $35\,{\rm T}$ a sharp metamagnetic transition occurs.~\cite{Sug99,Sch12}
The present results imply that the reconstruction (modification) of FS occur far below the metamagnetic transition field.
Its field response is dependent on the Fermi surface with different volume and effective mass.

The cyclotron effective mass also shows the field dependent behavior, as shown in Fig.~\ref{fig:Hdep_Freq_mass}(d).
The mass of branch $\beta$ rapidly decreases above $13\,{\rm T}$,
while branch $\alpha$ shows the increase of the mass approximately above $H^\ast=24\,{\rm T}$.
The mass of branch $\gamma$ also exhibits the gradual increase with field.
The new detected branch $\omega$ indicate the large mass ($25\,m_0$) in the limited measured field range.

\section{Summary}
We measured the SdH and dHvA oscillations using a high quality single crystal of URu$_2$Si$_2$.
Angular dependence of the SdH frequency confirmed that main branches consist of $\alpha$, $\beta$ and $\gamma$
originating from closed Fermi surfaces.
However, there is still a missing Fermi surface.
The unusual split of branch $\alpha$ was detected when the field is applied along the basal plane,
implying the formation of multi-domain.
The field dependent frequencies and the cyclotron effective masses are observed for all the main branches.
At high fields above $25\,{\rm T}$, a new branch named $\omega$ was detected.
These results indicate the reconstruction (or modification) of Fermi surfaces due to the magnetic field,
where the field-response strongly depends on the volume and effective mass of Fermi surface.
The small Fermi surfaces with heavy mass is more easily affected by the field as it is expected from the Zeeman energy consideration.

\section*{Acknowledgements}
We thank H. Harima, M. T. Suzuki, H. Kusunose, H. Ikeda, S. Kambe and W. Knafo for useful discussions.
This work was supported by ERC starting grant (NewHeavyFermion), French ANR project (CORMAT, SINUS) and EuromagNET II.


\begin{thebibliography}{10}
\expandafter\ifx\csname url\endcsname\relax
  \def\url#1{\texttt{#1}}\fi
\expandafter\ifx\csname urlprefix\endcsname\relax\def\urlprefix{URL }\fi

\bibitem{Sho84}
D.~Shoenberg: {\em Magnetic oscillations in metals} (Cambridge University
  Press, 1984).

\bibitem{Onu95a}
Y.~\={O}nuki and A.~Hasegawa: {\em Handbook on the Physics and Chemistry of
  Rare Earths}, eds. K.~A. {Gschneidner Jr.} and L.~Eyring (North-Holland,
  Amsterdam, 1995) Vol.~20, Chap.~135, p.~1.

\bibitem{Lif60}
I.~M. Lifshitz: Sov. Phys. JETP {\bf 11} (1960) 1130.

\bibitem{Bro87}
C.~Broholm, J.~K. Kjems, W.~J.~L. Buyers, P.~Matthews, T.~T.~M. Palstra, A.~A.
  Menovsky and J.~A. Mydosh: Phys. Rev. Lett. {\bf 58} (1987) 1467.

\bibitem{Bou03}
F.~Bourdarot, B.~F\aa{}k, K.~Habicht and K.~Proke{\v{s}}: Phys. Rev. Lett. {\bf
  90} (2003) 067203.

\bibitem{Wie07}
C.~R. Wiebe, J.~A. Janik, G.~J. Macdougall, G.~M. Luke, J.~D. Garrett, H.~D.
  Zhou, Y.~J. Jo, L.~Balicas, Y.~Qiu, J.~R.~D. Copley, Z.~Yamani and W.~J.~L.
  Buyers: Nature Physics {\bf 3} (2007) 96.

\bibitem{Vil08}
A.~Villaume, F.~Bourdarot, E.~Hassinger, S.~Raymond, V.~Taufour, D.~Aoki and
  J.~Flouquet: Phys. Rev. B {\bf 78} (2008) 012504.

\bibitem{Ohk99}
H.~Ohkuni, Y.~Inada, Y.~Tokiwa, K.~Sakurai, R.~Settai, T.~Honma, Y.~Haga,
  E.~Yamamoto, Y.~\={O}nuki, H.~Yamagami, S.~Takahashi and T.~Yanagisawa:
  Philos. Mag. B {\bf 79} (1999) 1045.

\bibitem{Bri95}
J.~P. Brison, N.~Keller, A.~Verni\`{e}re, P.~Lejay, L.~S.~A. Buzdin,
  J.~Flouquet, S.~R. Julian and G.~G. Lonzarich: Physica C {\bf 250} (1995)
  128.

\bibitem{Oka11}
R.~Okazaki, T.~Shibauchi, H.~J. Shi, Y.~Haga, T.~D. Matsuda, E.~Yamamoto,
  Y.~\={O}nuki, H.~Ikeda and Y.~Matsuda: Science {\bf 331} (2011) 439.

\bibitem{Has10_URu2Si2}
E.~Hassinger, G.~Knebel, T.~D. Matsuda, D.~Aoki, V.~Taufour and J.~Flouquet:
  Phys. Rev. Lett. {\bf 105}~(21) (2010) 216409.
  
\bibitem{Opp10}
P.~M. Oppeneer, J.~Rusz, S.~Elgazzar, M.-T. Suzuki, T.~Durakiewicz and J.~A.
  Mydosh: Phys. Rev. B {\bf 82} (2010) 205103.

\bibitem{Harima_pub}
H.~Harima: private communication.

\bibitem{Suzuki_pub}
M.~T. Suzuki: private communication.

\bibitem{Nak03} 
M.~Nakashima, H.~Ohkuni, Y.~Inada, R.~Settai, Y.~Haga, E.~Yamamoto and Y. ~\={O}nuki: 
J. Phys.: Condens. Matter {\bf 15} S2011 (2003).


\bibitem{Aok10}
D.~Aoki, F.~Bourdarot, E.~Hassinger, G.~Knebel, A.~Miyake, S.~Raymond,
  V.~Taufour and J.~Flouquet: J. Phys.: Condens. Matter {\bf 22} (2010)
  164205.

\bibitem{Mat11} 
T.~D. Matsuda, E. Hassinger, D. Aoki, V. Taufour, G. Knebel, N. Tateiwa, E. Yamamoto, Y. Haga, Y. \={O}nuki, Z. Fisk and J. Flouquet: 
J. Phys. Soc. Jpn. {\bf 80} (2011) 114710.

\bibitem{Onu99}
Y.~\={O}nuki, Y.~Inada, M.~Hedo, H.~Ohkuni, K.~Sakurai, Y.~Tokiwa, E.~Yamamoto,
  Y.~Haga, T.~Honma and S.~Takahashi: Physica B {\bf 259- 261} (1999) 1060.
  
\bibitem{Jo2007}
Y.~J.~Jo, L.~Balicas, C.~Capan, K.~Behnia, P.~Lejay, J.~Flouquet, J.~A.~Mydosh and P.~Schlottmann: Phys.~Rev.~Lett. {\bf 98} (2007) 166404.

\bibitem{Shi09}
H.~Shishido, K.~Hashimoto, T.~Shibauchi, T.~Sasaki, H.~Oizumi, N.~Kobayashi,
  T.~Takamasu, K.~Takehana, Y.~Imanaka, T.~D. Matsuda, Y.~Haga, Y.~Onuki and
  Y.~Matsuda: Phys. Rev. Lett. {\bf 102} (2009) 156403.

\bibitem{Altarawneh2011}
M.~M.~Altarawneh, N.~Harrison, S.~E.~Sebastian, L.~Balicas, P.~H.~Tobash, J.~D.~Thompson, F.~Ronning and E.~D.~Bauer:
Phys. Rev. Lett. {\bf 106} (2011) 146403.

\bibitem{Sch12} 
G. W. Scheerer, W. Knafo, D. Aoki, G. Ballon, A. Mari, D. Vignolles and J. Flouquet: 
Phys. Rev. B {\bf 85} 094402 (2012).

\bibitem{Elg09}
S.~Elgazzar, J.~Rusz, M.~Amft, P.~M. Oppeneer and J.~A. Mydosh: Nature
  Materials {\bf 8} (2009) 337.

\bibitem{Har09}
{H. Harima: private communication}.

\bibitem{Kas07}
Y.~Kasahara, T.~Iwasawa, H.~Shishido, T.~Shibauchi, K.~Behnia, Y.~Haga, T.~D.
  Matsuda, Y.~Onuki, M.~Sigrist and Y.~Matsuda: Phys. Rev. Lett. {\bf 99}
  (2007) 116402.

\bibitem{Gor06}
L.~P. Gor'kov and P.~D. Grigoriev: Phys. Rev. B {\bf 73}~(6) (2006) 060401.

\bibitem{Cra87}
G.~W. Crabtree, H.~Aoki, W.~Joss and F.~Hulliger: {\em Theoretical and
  Experimental Aspects of Valence Fluctuations and Heavy Fermions} (Plenum
  Press, New York, 1987) p.~197.

\bibitem{Ike12}
H.~Ikeda, M.~T. Suzuki, R.~Arita, T.~Takimoto, T.~Shibauchi and Y.~Matsuda:
  submitted.
  
\bibitem{Ter08}
T. Terashima, M. Kimata, S. Uji, T. Sugawara, N. Kimura, H. Aoki and H. Harima: Phys. Rev. B {\bf 78} (2008) 205107.

\bibitem{Malone2011}
L.~Malone, T.~D.~Matusda, A.~Antunes,  G.~Knebel,  V.~Taufour,  D.~Aoki, K.~Behnia, C.~Proust and J.~Flouquet: 
Phys.~Rev.~B {\bf 83} (2011) 245117.

\bibitem{Taka96}
M.~Takashita, H.~Aoki, T.~Terashima, S.~Uji, K.~Maezawa, R.~Settai and
  Y.~\={O}nuki: J. Phys. Soc. Jpn. {\bf 65} (1996) 515.

\bibitem{She03}
I.~Sheikin, A.~{Gr{\"o}ger}, S.~Raymond, D.~Jaccard, D.~Aoki, H.~Harima and
  J.~Flouquet: Phys. Rev. B {\bf 67} (2003) 094420.

\bibitem{Mcc05}
A.~{McCollam}, S.~R. Julian, P.~M.~C. Rourke, D.~Aoki and J.~Flouquet: Phys.
  Rev. Lett. {\bf 94} (2005) 186401.

\bibitem{Kim00}
N.~Kimura, M.~Nakayama, S.~Uji, D.~Aoki, Y.~Haga, E.~Yamamoto, Y.~\={O}nuki and
  H.~Aoki: Physica B {\bf 284-288} (2000) 1279.

\bibitem{Sug99}
K.~Sugiyama, M.~Nakashima, H.~Ohkuni, K.~Kindo, Y.~Haga, T.~Honma, E.~Yamamoto
  and Y.~\={O}nuki: J. Phys. Soc. Jpn. {\bf 68}~(10) (1999) 3394.


\end{thebibliography}

\end{document}